\documentclass[showpacs,aps,prl,floatfix,reprint,superscriptaddress,nofootinbib]{revtex4-1}
\usepackage{xfrac}
\usepackage{amssymb,amsfonts,amsmath}
\usepackage{braket}
\usepackage{graphicx}
\usepackage{verbatim}
\usepackage{etoolbox}
\usepackage[utf8]{inputenc} %useful to type directly diacritic characters
\usepackage{mathtools}
\usepackage{soul,xcolor,colortbl}
\usepackage{setspace}
\usepackage{multirow}
\usepackage{makecell}
\usepackage[normalem]{ulem}
\usepackage{scrextend}

\usepackage{hyperref} \hypersetup{colorlinks=true,citecolor=blue,linkcolor=blue,urlcolor=blue} % HYPERLINKS

\definecolor{pranab_green}{rgb}{0.31,0.53,0.10}
\definecolor{pranab_red}{rgb}{0.85,0.23,0.11}

\usepackage{bm} %revtex4.1 for bold symbols
\usepackage{array}
\newcolumntype{C}[1]{>{\centering\arraybackslash}p{#1}}\usepackage{soul}
\definecolor{Gray}{gray}{0.85}

\usepackage{color} % STEFANO
\usepackage{morefloats}

\definecolor{Gray}{gray}{0.9}
\definecolor{LightCyan}{rgb}{0.88,1,1}

\def\AGL{{\small AGL}}
\def\AEL{{\small AEL}}
\def\VRH{{\small VRH}}

\def\ROM{{\small ROM}}
\def\fROM{{\footnotesize ROM}}
\def\AFLOW{{\small AFLOW}}
\def\fAFLOW{{\footnotesize AFLOW}}
\def\AFLOWPOCC{{\small AFLOW-POCC}} 
\def\AFLOWCHULL{{\small AFLOW-CHULL}} 

 \def\EFA{{\small EFA}} % AFLOW-TEX COMMON

\def\XRD{{\small XRD}}
\def\SPS{{\small SPS}}
\def\FCC{fcc}
\def\CALPHAD{{\small CALPHAD}}    % AFLOW-TEX COMMON
\def\LTVC{{\small LTVC}}    % AFLOW-TEX COMMON

\makeatletter \renewcommand\frontmatter@abstractwidth{\dimexpr\textwidth\relax} \makeatother % WIDEABSTRACT

\begin{document}
% \title{Novel high-entropy carbides discovered by synthesizability descriptors}
\title{High-entropy high-hardness metal carbides discovered by entropy descriptors}

\author{Pranab Sarker}
\thanks{These authors contributed equally to the work}
\affiliation{Department of Mechanical Engineering and Materials Science, Duke University, Durham, NC 27708, USA}
\author{Tyler Harrington}
\thanks{These authors contributed equally to the work}
\affiliation{Materials Science and Engineering Program, University of California, San Diego, La Jolla, CA 92093, USA}
\author{Cormac Toher}
\affiliation{Department of Mechanical Engineering and Materials Science, Duke University, Durham, NC 27708, USA}
\author{Corey Oses}
\affiliation{Department of Mechanical Engineering and Materials Science, Duke University, Durham, NC 27708, USA}
\author{Mojtaba Samiee}
\affiliation{Department of NanoEngineering, University of California, San Diego, La Jolla, CA 92093, USA}
\author{\\Jon-Paul Maria}
\affiliation{Department of Materials Science and Engineering, North Carolina State University, Raleigh, NC 27695, USA}
\author{Donald W. Brenner}
\affiliation{Department of Materials Science and Engineering, North Carolina State University, Raleigh, NC 27695, USA}
\author{Kenneth S. Vecchio}
\email[]{kvecchio@eng.ucsd.edu}
\affiliation{Materials Science and Engineering Program, University of California, San Diego, La Jolla, CA 92093, USA}
\affiliation{Department of NanoEngineering, University of California, San Diego, La Jolla, CA 92093, USA}
\author{Stefano Curtarolo}
\email[]{stefano@duke.edu}
\affiliation{Materials Science, Electrical Engineering, Physics and Chemistry, Duke University, Durham NC, 27708, USA}
\affiliation{Fritz-Haber-Institut der Max-Planck-Gesellschaft, 14195 Berlin-Dahlem, Germany}

\date{\today}

\begin{abstract}
  \noindent
  % {\bf ABSTRACT.}
  High-entropy materials have attracted considerable interest due to the combination of useful properties and promising applications. 
  Predicting their formation remains the major hindrance to the discovery of new systems. 
  Here we propose a descriptor --- entropy forming ability --- for addressing synthesizability from first principles. 
  The formalism, based on the energy distribution spectrum of randomized calculations, 
  captures the accessibility of equally-sampled states near the ground state and quantifies configurational disorder capable of stabilizing high-entropy homogeneous phases. 
  The methodology is applied to disordered refractory 5-metal carbides --- promising candidates for high-hardness applications. 
  The descriptor correctly predicts the ease with which compositions can be experimentally synthesized as rock-salt high-entropy homogeneous phases, validating the ansatz, and in some cases, going beyond intuition. 
  Several of these materials exhibit hardness up to 50\% higher than rule of mixtures estimations. 
  The entropy descriptor method has the potential to accelerate the search for high-entropy systems by rationally combining first principles with experimental synthesis and characterization. 
\end{abstract}
\maketitle
% \pacs{66.70.-f, 66.70.Df}

{\section*{Introduction}}

High-entropy materials having a highly disordered homogeneous crystalline single-phase (potentially stabilized entirely by entropic contributions)
continue to attract a great deal of research interest \cite{Gao_HEA_book_2015, Miracle_HEAs_NComm_2015, Widom_HEA_JMR_2018}.
{Remarkable properties have been reported: high 
  % melting point, 
  strength (yield stress $> 1$GPa) combined with ductility  \cite{HEAapp1,Ye_hea_high_melting_point,Gludovatz_hea_mech_properties,Gali_hea_mech_properties,MoNbTaW2,HEAprop2,Tsao_HESA_SciRep_2017,Li_SRep_HEA_Dual_Phase_2017}, 
  hardness \cite{Ye_hea_high_melting_point,MoNbTaW,Senkov_HEA_AM_2013}, superconductivity \cite{HEAprop1}, colossal dielectric constant \cite{Beradan_2016_PSSA_ESO_Colossal}, and superionic conductivity \cite{Beradan_2016_JMCA_ESO_superionic}.}
{Entropy is thought to play a key-stabilizing role in high-entropy alloys \cite{Gao_HEA_book_2015}, 
  entropy-stabilized oxides \cite{curtarolo:art99, curtarolo:art122}, high-entropy borides \cite{Gild_borides_SciRep_2016} 
  and high-entropy carbides \cite{Castle_2018_4metalC, Dusza_2018_4metalC, Yan_2018_5metalC_thermal_conductivity, Zhou_2018_5metalC_oxidation}.
  The latter three classes consist of disordered metal cation sublattices with several species at equi-concentration
  combined with oxide \cite{curtarolo:art99,curtarolo:art122,Beradan_2016_PSSA_ESO_Colossal,Heron_2017_SCIREP_Exchange_Coupling}, boride \cite{Gild_borides_SciRep_2016}
  or carbide \cite{Castle_2018_4metalC, Dusza_2018_4metalC, Yan_2018_5metalC_thermal_conductivity, Zhou_2018_5metalC_oxidation} anion sublattices.} 
These systems offer the potential to combine excellent thermo-mechanical properties and resilient thermodynamic stability given by entropy stabilization with the higher oxidation resistance of ceramics \cite{curtarolo:art80}.
% promising candidates for ultra-high-temperature applications \cite{Wuchina_UHTCs_borides_carbides_nitrides_2007}.
% \st{Disordered carbides can be considered for ultra-high-temperature} \cite{Wuchina_UHTCs_borides_carbides_nitrides_2007} \st{and high-hardness applications.}
{The resistance of disordered carbides to extreme heat \cite{Agte_Carbides_ZtP_1930, Andrievskii_Carbides_PM_1967, Hong_HfTaC_PRB_2015}, oxidation \cite{Zhou_2018_5metalC_oxidation} and wear
  makes them promising ultra-high-temperature ceramics for thermal protection coatings in aerospace applications \cite{Wuchina_UHTCs_borides_carbides_nitrides_2007},
  and as high-hardness, relatively low density high-performance drill bits and cutting tools in mining and industry.}

Super-hard transition metal carbides have been known since the 1930s to exhibit significant levels of solid solution \cite{Agte_Carbides_ZtP_1930, Rudy_TernaryPhases2_1965, Rudy_TernaryPhases5_1969}, 
and to display high melting temperatures \cite{Agte_Carbides_ZtP_1930, Andrievskii_Carbides_PM_1967}.
{Ta$_x$Hf$_{1-x}$C forms a homogeneous solid solution across all composition ranges \cite{Gusev_Carbides_RJPC_1985, Hong_HfTaC_PRB_2015, CedillosBarraza_HfTaC_JECS_2016}, 
  with Ta$_4$HfC$_5$ exhibiting one of the highest experimentally reported melting points ($T_{\rm m} \sim 4263$K \cite{Agte_Carbides_ZtP_1930, Andrievskii_Carbides_PM_1967}). 
  In this case, the two refractory metals randomly populate one of the two rock-salt sublattices \cite{Hong_HfTaC_PRB_2015}.}
% and {amongst} the
% highest experimentally reported melting point ($T_{\rm m} \sim 4263$K \cite{Agte_Carbides_ZtP_1930, Andrievskii_Carbides_PM_1967}) 
% occurs for Ta$_4$HfC$_5$, 
% in which the two refractory metals randomly populate one of the two rock-salt sublattices \cite{Hong_HfTaC_PRB_2015}. 
{More recent measurements indicate that the maximum melting point of $4232$K occurs without Ta at the composition HfC$_{0.98}$ \cite{CedillosBarraza_HfTaC_SciRep_2016}.}
{Investigation of new carbide compositions will help elucidate the high temperature behavior of these materials, and will provide an avenue to settle the discrepancies in the experimental literature.}
% {To find novel materials with even more advantageous properties, including increased thermal stability, enhanced strength and hardness,  %CO181009
{To {discover} materials with even more advantageous properties, including increased thermal stability, enhanced strength and hardness,
  and improved oxidation resistance \cite{Zhou_2018_5metalC_oxidation}, more species and configurations have to be considered.}
Unfortunately, the lack of a rational, effective and rapid method to find and characterize the disordered crystalline phase
makes it impossible to pinpoint the right combination of species/compositions and the discovery continues by slow
and relatively expensive trial and error.

Computationally, the hindrance in {\it in-silico} disordered materials development can be attributed to entropy --- a very difficult quantity to parameterize 
when searching through the immense space of candidates 
% (even with efficient computational methods, {\it e.g.} Monte Carlo and {\it ab-initio} lattice energies in the Wang-Landau \cite{WangLandau_AJP2004} % \cite{WangLandau_PRL2001, WangLandau_PRE2001, WangLandau_AJP2004} %CO181009
(even with efficient computational methods, {\it e.g.} Monte Carlo and {\it ab-initio} lattice energies in the Wang-Landau \cite{WangLandau_AJP2004} % \cite{WangLandau_PRL2001, WangLandau_PRE2001, WangLandau_AJP2004} 
or nested sampling \cite{Csanyi_NestedSampling_PRB_2016} formalisms).
{\CALPHAD\ has also been applied successfully \cite{CoFeMnNi,Zhang_HEA_Calphad_2014,Miracle_HEAs_NComm_2015,Gao_HEA_Opinion_2017,Senkov_HEA_AM_2013}, 
  although it is dependent on the availability of sufficient experimental data.}
This is the perfect challenge for {\it ab-initio} high-throughput computing \cite{nmatHT}
% as long as reasonable entropy {\it descriptors} --- the set of parameters capturing the underlying mechanism of a materials property --- can be found.
as long as reasonable entropy {descriptors} --- the set of parameters capturing the underlying mechanism of a materials property --- can be found.
% In this article, we undertake the challenge by formulating an ``{\it \underline{e}ntropy-\underline{f}orming-\underline{a}bility}'' descriptor (\EFA). %CO181009

In this article, we undertake the challenge by formulating an {\underline{e}ntropy-\underline{f}orming-\underline{a}bility} descriptor (\EFA).
{It captures the relative propensity of a material %at a given composition
  to form a high-entropy single-phase crystal
  by measuring the energy distribution (spectrum) of configurationally-randomized calculations up to a given unit-cell size.}
A narrow spectrum implies a low energy cost for accessing metastable configurations, % for a given composition,
hence promoting randomness %of the cations 
(\textit{i.e.} entropy) into the system {(high-\EFA)} at finite temperature.
In contrast, a wide spectrum suggests a composition with a high energy
barrier for introducing different configurations {(low-\EFA)}, and thus with a strong preference for ordered phases. 
The method is benchmarked by the matrix of possible carbides.
Given a set of 8 refractory metals %(Nb, Ta, Mo, W, Ti, Zr, Hf, V) 
{(Hf, Nb, Mo, Ta, Ti, V, W, Zr)} plus carbon, the formalism predicts the matrix of synthesizable 5-metal high-entropy carbides.
Candidates are then experimentally prepared, leading to a novel class of systems.
{In particular, it is demonstrated that the descriptor is capable of reliably distinguishing between the compositions that easily form homogeneous single phases and the ones that do not,
  including identifying compositions that form single phases despite incorporating multiple binary carbide precursors with different structures and stoichiometric ratios.}
{Note that because of the differing stoichiometries of the non-rock-salt phase binary carbide precursors Mo$_2$C and W$_2$C, the compositions listed in this work are nominal. 
  There are extensive carbon vacancies in the anion sublattice in the synthesized materials, further contributing to the configurational entropy.}
Several of these materials display enhanced mechanical properties, {\it e.g.} Vickers hardness up to 50\% higher than predicted by a rule of mixtures. 
Thus, this class of materials has strong potential for industrial
uses where dense and wear resistant impactors are needed, particularly for extreme temperature applications.
The successful outcome demonstrates the strength of the synergy
between thermodynamics, high-throughput computation, and experimental
synthesis.

{\small
  \begin{table*}
    \caption{\small {Results for the calculated entropy forming ability (\EFA) descriptor, energetic distance from 6-dimensional convex hull ($\Delta H_{\rm f}$) 
        and vibrational free energy at 2000K ($\Delta F_{\rm vib}$) for the 5-metal carbide systems, arranged in descending order of \EFA.}     
      9 compositions are selected for experimental investigation. 
      The lattice distortion, $\varepsilon$, is obtained from the peak broadening in XRD. 
      % ``S'': single-phase formed; ``M'': multi-phase formed in experiment.   %CO181009
      {S: single-phase formed; M: multi-phase formed in experiment.}
      {Note that the compositions listed here are nominal, and the actual synthesized compositions can vary due to the presence of carbon vacancies in the anion sublattice.}
      Units: \EFA\ in $\mathrm{(eV/atom)}^{-1}$; $\Delta H_{\rm f}$ and $\Delta F_{\rm vib}$ in $\mathrm {(meV/atom)}$; and $\varepsilon$ in \%.}
    \label{table1}
    % \footnotesize 
    \resizebox{\textwidth}{!} {
      % \begin{adjustbox}{width=1.2\textwidth,center=\textwidth}
      \begin{tabular}{||l|c|c|c|c|c||l|c|c|c|c|c||l|c|c|c|c|c||}
        \hline
        system &	EFA & $\Delta H_{\rm f}$ & $\Delta F_{\rm vib}$	& exp. &	$\varepsilon$	& system	& EFA & $\Delta H_{\rm f}$ & $\Delta F_{\rm vib}$ &	exp.	& $\varepsilon$ & system & EFA &	$\Delta H_{\rm f}$ &	$\Delta F_{\rm vib}$	& exp. &	$\varepsilon$	\\ \hline
        MoNbTaVWC$_{5}$	&	125	&	156	&	-14	&	S	&	0.063	&	HfNbTaVWC$_{5}$	&	67	&	110	&		&		&		&	TaTiVWZrC$_{5}$	&	50	&	96	&		&		&		\\ \hline
        HfNbTaTiZrC$_{5}$	&	100	&	19	&	-12	&	S	&	0.094	&	HfMoTaTiVC$_{5}$	&	67	&	82	&		&		&		&	NbTiVWZrC$_{5}$	&	50	&	93	&		&		&		\\ \hline
        HfNbTaTiVC$_{5}$	&	100	&	56	&	-31	&	S	&	0.107	&	HfMoNbTiZrC$_{5}$	&	67	&	53	&		&		&		&	HfMoTiVZrC$_{5}$	&	50	&	96	&		&		&		\\ \hline
        MoNbTaTiVC$_{5}$	&	100	&	82	&		&		&		&	MoNbTaWZrC$_{5}$	&	63	&	133	&		&		&		&	HfMoTaVZrC$_{5}$	&	50	&	92	&		&		&		\\ \hline
        NbTaTiVZrC$_{5}$	&	83	&	64	&		&		&		&	HfMoTaTiZrC$_{5}$	&	63	&	55	&		&		&		&	HfMoNbVZrC$_{5}$	&	50	&	89	&		&		&		\\ \hline
        HfMoNbTaTiC$_{5}$	&	83	&	48	&		&		&		&	NbTaTiWZrC$_{5}$	&	59	&	61	&		&		&		&	MoTaVWZrC$_{5}$	&	48	&	148	&		&		&		\\ \hline
        NbTaTiVWC$_{5}$	&	77	&	92	&	-19	&	S	&	0.124	&	MoTaTiVZrC$_{5}$	&	59	&	92	&		&		&		&	MoTaTiWZrC$_{5}$	&	48	&	94	&		&		&		\\ \hline
        MoNbTaTiWC$_{5}$	&	77	&	111	&		&		&		&	MoNbTiVZrC$_{5}$	&	59	&	87	&		&		&		&	MoNbVWZrC$_{5}$	&	48	&	146	&		&		&		\\ \hline
        MoNbTiVWC$_{5}$	&	71	&	122	&		&		&		&	MoNbTaVZrC$_{5}$	&	59	&	108	&		&		&		&	MoNbTiWZrC$_{5}$	&	48	&	89	&		&		&		\\ \hline
        MoNbTaTiZrC$_{5}$	&	71	&	57	&		&		&		&	HfNbTiVWC$_{5}$	&	59	&	81	&		&		&		&	HfMoNbWZrC$_{5}$	&	48	&	101	&		&		&		\\ \hline
        HfTaTiVZrC$_{5}$	&	71	&	73	&		&		&		&	HfNbTaWZrC$_{5}$	&	59	&	53	&		&		&		&	HfTiVWZrC$_{5}$	&	45	&	99	&		&		&		\\ \hline
        HfNbTiVZrC$_{5}$	&	71	&	73	&		&		&		&	NbTaVWZrC$_{5}$	&	56	&	119	&		&		&		&	HfNbVWZrC$_{5}$	&	45	&	94	&		&		&		\\ \hline
        HfMoNbTiVC$_{5}$	&	71	&	77	&		&		&		&	HfTaTiVWC$_{5}$	&	56	&	84	&		&		&		&	HfMoTiVWC$_{5}$	&	45	&	97	&		&		&		\\ \hline
        HfMoNbTaZrC$_{5}$	&	71	&	48	&		&		&		&	HfMoTaVWC$_{5}$	&	56	&	139	&		&		&		&	HfMoTaWZrC$_{5}$	&	45	&	105	&		&	M	&	0.271	\\ \hline
        HfMoNbTaWC$_{5}$	&	71	&	126	&		&		&		&	HfMoNbVWC$_{5}$	&	56	&	137	&		&		&		&	HfTaVWZrC$_{5}$	&	43	&	97	&		&		&		\\ \hline
        HfMoNbTaVC$_{5}$	&	71	&	99	&		&		&		&	HfNbTiWZrC$_{5}$	&	53	&	56	&		&		&		&	MoTiVWZrC$_{5}$	&	40	&	107	&		&		&		\\ \hline
        HfNbTaTiWC$_{5}$	&	67	&	53	&$\sim0$&	S	&	0.171	&	HfMoTaTiWC$_{5}$	&	53	&	84	&		&		&		&	HfMoTiWZrC$_{5}$	&	38	&	83	&		&	M	&	0.315	\\ \hline
        MoTaTiVWC$_{5}$	&	67	&	128	&		&		&		&	HfMoNbTiWC$_{5}$	&	53	&	81	&		&		&		&	HfMoVWZrC$_{5}$	&	37	&	141	&		&	M	&	0.325	\\ \hline
        HfNbTaVZrC$_{5}$	&	67	&	60	&		&		&		&	HfTaTiWZrC$_{5}$	&	50	&	59	&$\sim0$&	S	&	0.169	&		&		&		&		&		&		\\ \hline
      \end{tabular}
      % \end{adjustbox}
    }
  \end{table*}
}

{\section*{Results}}

\noindent
{\bf \EFA\ formalism.}
To accelerate the search in the chemical space, the entropy content of a
compound is estimated from the energy distribution spectrum of metastable configurations above the zero-temperature ground-state.
At finite $T$, any disordered state can be present with a probability given by the Boltzmann distribution and the state's degeneracy.
Note that configurations are randomly sampled up to a given unit-cell size: 
the larger the size, the more accurately the spectrum represents the real thermodynamic density of states.

{The energy distribution ($H_{i}$) spectrum can be
  quantitatively characterized by its standard deviation $\sigma$, 
  so that the $\sigma$  becomes the descriptor for $S$: the smaller $\sigma$, the larger $S$.
  % The descriptor for a $N$-species system, called the ``{\it \underline{e}ntropy \underline{f}orming \underline{a}bility}'' (\EFA),  %CO181009
  The descriptor for an $N$-species system, called the {\underline{e}ntropy \underline{f}orming \underline{a}bility} (\EFA),
  is defined as the inverse of the $\sigma$ of the energy spectrum above 
  the ground-state of the $N$-system at zero temperature: 
  \begin{equation}
    % {\rm EFA}(N)\equiv \left\{\sigma \left[ {\rm spectrum}(H_i(N))\right]_{\small T=0} \right\}^{-1},  %CO181010
    {\rm EFA}(N)\equiv \left\{\sigma \left[ {\rm spectrum}(H_i(N))\right]_{T=0} \right\}^{-1},
    \vspace{-2mm}
    \label{EFA1}
  \end{equation} }
\noindent {where
  \begin{equation}
    \sigma \left\{ H_i(N)\right\}= \sqrt {\frac{\sum\limits_{i =1}^{n}g_{i}(H_{i}-H_{\rm mix})^{2}}{\left(\sum\limits_{i=1}^{n} g_{i}\right) - 1}},
    \vspace{1 mm}
    \label{sigma}
  \end{equation} 
  \noindent  where $n$ is the total number of sampled geometrical configurations and $g_{i}$ are their degeneracies. 
  $H_{\rm mix}$ is the mixed-phase enthalpy approximated by averaging the enthalpies $H_{i}$ of the sampled configurations:
  \begin{equation}
    H_{\rm mix} = \frac{\sum\limits_{i=1}^{n}g_{i}H_{i}}{\sum\limits_{i=1}^{n}g_{i}}.
    \vspace{1 mm}
    \label{Hmix}
  \end{equation} 
  \noindent The broader the spectrum, the more energetically expensive it will be to introduce configurational disorder into the system, and thus
  the lower the \EFA. \EFA\ is measured in $\mathrm{(eV/atom)}^{-1}$}.
\begin{figure*}
  \centerline{\includegraphics[width=0.98\textwidth]{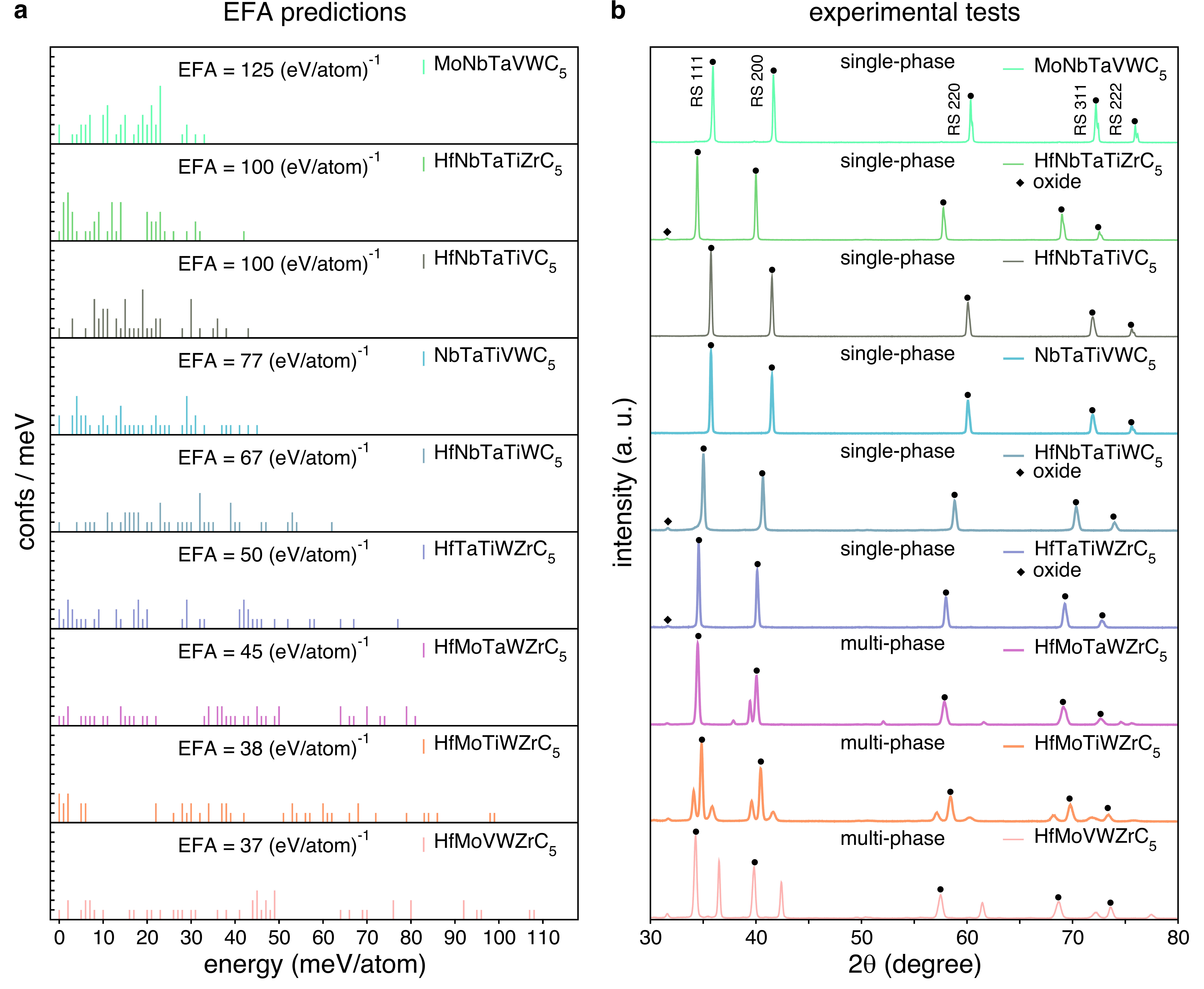}} 
  % \centerline{\includegraphics[width=0.99\textwidth]{test.pdf}} 
  \vspace{-2mm}
  \caption{\small
    {\bf Schematics of high-entropy carbides predictions.}
    {\bf (a)} The energy distribution of different configurations of the 9 5-metal carbides: 
    {MoNbTaVWC$_5$}, HfNbTaTiZrC$_5$, HfNbTaTiVC$_5$, NbTaTiVWC$_5$, HfNbTaTiWC$_5$, HfTaTiWZrC$_5$, HfMoTaWZrC$_5$, HfMoTiWZrC$_5$, and HfMoVWZrC$_5$;
    {spectrum is shifted so that the lowest energy configuration for each composition is at zero.
      The energy spectrum for each composition indicates its propensity to form the high-entropy single-phase:
      the narrower the distribution, the more likely it is to form a high-entropy single-phase at finite $T$}.  
    {\bf (b)} The x-ray diffraction patterns for the same 9 5-metal carbides, where the first 6 compositions exhibit only the 
    desired \FCC\ structure peaks, while the additional peaks for remaining 3 compositions indicate the presence of secondary phases.
    {The small peaks at $2\theta = 31.7^{\circ}$, marked by the diamond symbol ($\blacklozenge$) in the spectra of HfNbTaTiZrC$_5$, HfNbTaTiWC$_5$ and HfTaTiWZrC$_5$,
      are from the (111) plane of a monoclinic (Hf,Zr)O$_2$ phase that remains due to processing.}
  } 
  \label{fig1}
\end{figure*}

\begin{figure*}%[h!]
  % \centerline{\includegraphics[width=0.9\textwidth]{xrd_eds_hec.pdf}} 
  \centerline{\includegraphics[width=0.98\textwidth]{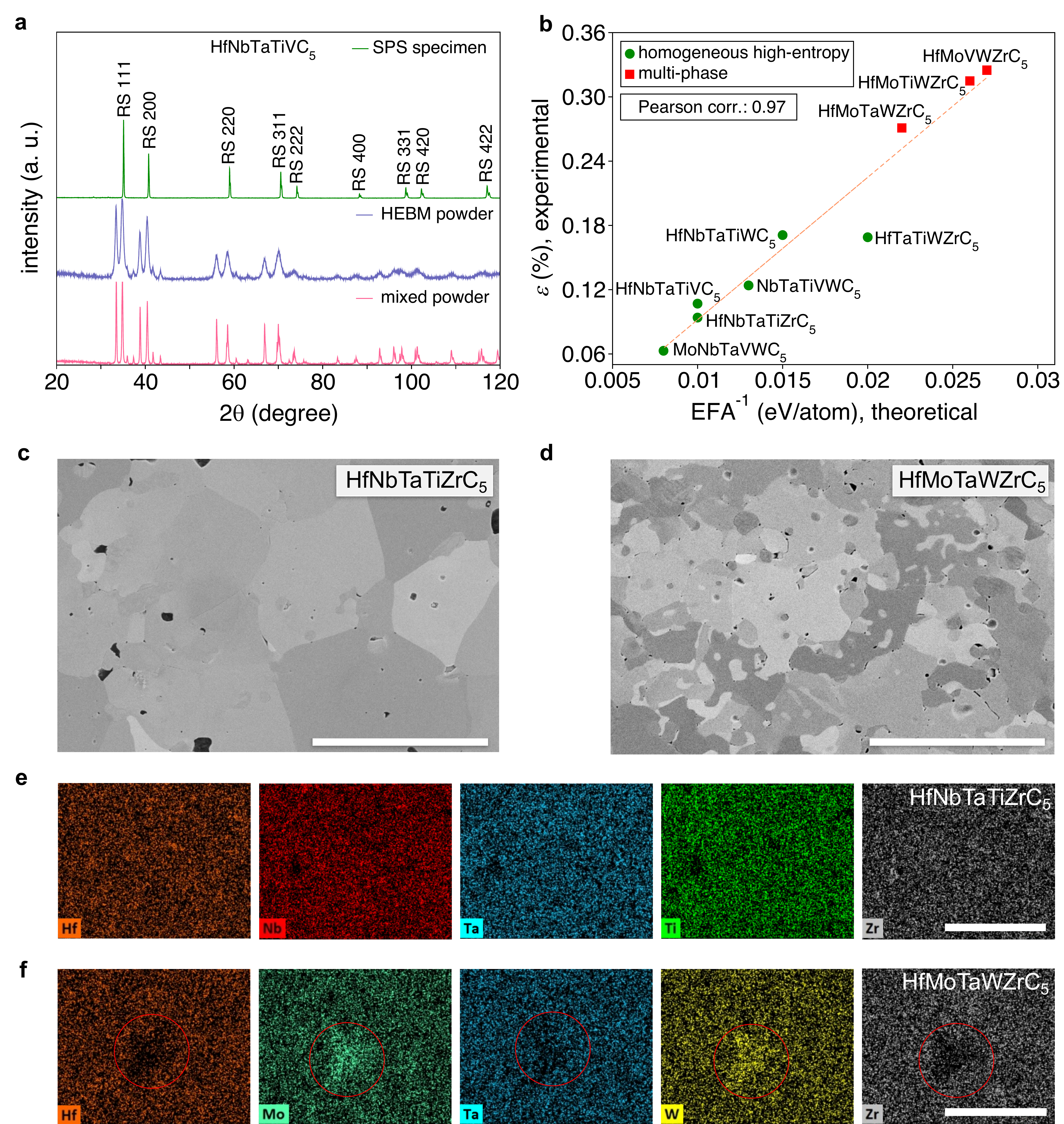}}
  \vspace{-3mm}
  \caption{\small {\bf Experimental results for high-entropy carbides synthesis and characterization.} 
    {\bf (a)} Progression of a sample of HfNbTaTiVC$_5$ through each processing step: hand mixing (magenta spectrum, bottom), ball milling (blue spectrum, center), 
    and spark plasma sintering (green spectrum, top), depicting the evolution towards the desired rock-salt crystal structure ($a_{\mathrm{exp}} = 4.42$\AA).
    {\bf (b)} {Linear relationship between \EFA{}$^{-1}$} and the distortion of experimental lattice parameters $\varepsilon$.
    {Green circles (\textcolor{pranab_green}{$\bullet$}) and red
      squares (\textcolor{pranab_red}{$\blacksquare$})} indicate homogeneous high-entropy single- and multi-phase compounds, respectively. 
    {\bf (c, d)} Electron micrographs of single-phase HfNbTaTiZrC$_5$ and multi-phase HfMoTaWZrC$_5$ specimens. 
    {\bf (e, f)} Selected EDS compositional maps of the HfNbTaTiZrC$_5$ and HfMoTaWZrC$_5$ 
    specimens. The micrographs show the presence of the secondary
    phase in HfMoTaWZrC$_5$ (circles) that is also present in \XRD\
    results, which is revealed to be a W- and Mo-rich phase. {Scale bars,
      $10\mu {\rm m}$ (c-f)}.
  } 
  \label{fig2}
\end{figure*}

\noindent
{\bf \EFA\ calculation.}
A total of 56 5-metal compositions can be generated using the 8 refractory metals ($8!/5!3! = 56$) of interest %(Nb, Ta, Mo, W, Ti, Zr, Hf, and V).
{(Hf, Nb, Mo, Ta, Ti, V, W, Zr).}
For each composition, the Hermite normal form superlattices of the 
\AFLOW\ \underline{p}artial \underline{occ}upation (\AFLOWPOCC) method \cite{curtarolo:art110} 
generate {49 distinct 10-atom-cell configurations, resulting in a total of 2744 configurations needed to determine the \EFA\ of this composition space (Methods section).}
% The \AFLOWPOCC\ method has advantages over other approaches in that it does not require large supercells, 
% and also avoids the implicit bias toward the disordered states with the lowest site correlations inherent in other methods such as special quasirandom structures \cite{zunger_sqs}.} 
{The {\it ab-initio} calculated \EFA\ values for the full set of 56 5-metal compositions are provided in Table \ref{table1}. 
  Nine candidates are chosen from this list for experimental synthesis and investigation: 
  \textbf{i.} the three candidates with the highest value of \EFA\ (MoNbTaVWC$_5$ (\EFA\ = $125~\mathrm{(eV/atom)}^{-1}$), HfNbTaTiZrC$_5$ (\EFA\ = $100~\mathrm{(eV/atom)}^{-1}$),  HfNbTaTiVC$_5$ (\EFA\ = $100~\mathrm{(eV/atom)}^{-1}$), 
  high probability of forming high-entropy single-phases),
  \textbf{ii.} the two candidates with the lowest value of \EFA\ (HfMoVWZrC$_5$ ($37~\mathrm{(eV/atom)}^{-1}$), HfMoTiWZrC$_{5}$ ($38~\mathrm{(eV/atom)}^{-1}$), low probability of forming high-entropy single-phases), 
  and \textbf{iii.} four chosen at random with intermediate \EFA\
  (NbTaTiVWC$_5$ ($77~\mathrm{(eV/atom)}^{-1}$), HfNbTaTiWC$_5$ ($67~\mathrm{(eV/atom)}^{-1}$), HfTaTiWZrC$_5$ ($50~\mathrm{(eV/atom)}^{-1}$), and HfMoTaWZrC$_5$ ($45~\mathrm{(eV/atom)}^{-1}$)).
  % The candidate with the lowest value of \EFA\ , HfMoVWZrC$_5$ ($37$~eV$^{-1}$), is also selected to probe the relative importance of \EFA\ 
  % for predicting the formation of high-entropy homogeneous single-phases.
  Figure \ref{fig1}(a) shows the energy distribution and \EFA\ values obtained {\it ab initio} from configurations generated with \AFLOW\ for the nine chosen systems. 
  For MoNbTaVWC$_5$, HfNbTaTiZrC$_5$, and HfNbTaTiVC$_5$, most of the configurations are within 20~meV/atom of the lowest energy state, 
  and the distributions have an \EFA\ of at least $100~\mathrm{(eV/atom)}^{-1}$. 
  Therefore, at finite temperature, most of the configurations should have a high probability of being formed, 
  so that a high level of configurational randomness is expected to be accessible in the three systems, 
  making them promising candidates to form a high-entropy homogeneous single-phase. 
  Achieving a similar level of configurational randomness would be progressively more difficult in NbTaTiVWC$_5$, HfNbTaTiWC$_5$,  and 
  HfTaTiWZrC$_5$ as the different configurations display a broader energy distribution, with \EFA{}s ranging from $50$ to $77~\mathrm{(eV/atom)}^{-1}$. 
  A higher energy cost is needed to incorporate configurational entropy into these 3 compositions, so forming a homogeneous 
  single-phase will be more difficult.
  For HfMoTaWZrC$_5$, HfMoTiWZrC$_5$, and HfMoVWZrC$_5$, the spread of energies for the configurations is very wide, 
  with \EFA\ values from $45$ down to $37~\mathrm{(eV/atom)}^{-1}$. 
  These materials would be expected to be very difficult to synthesize as a homogeneous single-phase.} 
% \noindent

\noindent
{{\bf Competing ordered phases.}
  The phase diagrams for the 5-metal carbide systems were generated to investigate the existence of binary and ternary ordered structures that could compete with the formation of the high entropy single phase. 
  First, prototypes for experimentally reported binary and ternary carbide structures \cite{ICSD, Massalski} are used to calculate the formation enthalpies of additional ordered phases for the \AFLOW\ database. 
  The results are used to generate the convex hull phase diagrams for
  all 56 compositions using the \AFLOWCHULL\ module
  \cite{Oses_CHULL_JCIM_2018}. 
  {The relevant binary and ternary
    convex  hulls are illustrated in Supplementary Figures 1-36.}
  The distance along the enthalpy axis of the lowest energy \AFLOWPOCC\ configuration from the convex hull, $\Delta H_{\mathrm f}$, is listed in Table \ref{table1} 
  (the decomposition reaction products are summarized in {Supplementary Table 2}). 
  A rough estimation of the synthesis temperature $T_{\rm s}$ (see the analogous Entropy Stabilized Oxides case, Fig. 2 of Ref. \onlinecite{curtarolo:art99}) 
  --- can be calculated by dividing $\Delta H_{\mathrm f}$ by the ideal configuration entropy ({Supplementary Table 1}, 
  % with the ideal entropy per atom evaluated as $0.5 k_{\rm B} \times \log 0.2$, since $k_{\rm B}\times \log 0.2$ is the entropy per metal-carbide atomic pair)\footnote{{A precise   %CO181009
  % characterization of the disorder requires more expensive} {approaches, such as the \LTVC\ method \cite{curtarolo:art139}, and is beyond the scope of this article.}}. %CO181009
  with the ideal entropy per atom evaluated as $0.5 k_{\rm B} \times \log 0.2$, since $k_{\rm B}\times \log 0.2$ is the entropy per metal-carbide atomic pair). 
  {{A precise characterization of the disorder requires more expensive} {approaches, such as the \LTVC\ method \cite{curtarolo:art139}, and is beyond the scope of this article.}}
  The highest temperature is 2254K, which is less than the synthesis temperature of $2200^{\circ}$C (2473K), 
  indicating that during sintering the disordered phases are thermodynamically accessible with respect to decomposition into ordered compounds. 
  % In analogy to what happens to metallic glasses 
  % In analogy to the formation of metallic glasses where energetic ``confusion'' obstructs crystalline growth \cite{curtarolo:art112, greer1993confusion} once the temperature is reduced,  %CO181009
  In analogy to the formation of metallic glasses where energetic {confusion} obstructs crystalline growth \cite{curtarolo:art112, greer1993confusion} once the temperature is reduced, 
  systems with high \EFA\ remain locked into ensembles of highly degenerate configurations, retaining the disorder achieved at high temperature. 
  Hence, \EFA\ provides a measure of the relative synthesizability of the disordered composition.}

\noindent
{{\bf From enthalpy to entropy.}
  It is important to consider that the metal-carbide precursors have very strong covalent/ionic bonds, and are therefore enthalpy stabilized \cite{Kuo_MoC_Nature_1952}. However, the same might not be the case for their mixture.  
  % In fact, a statistical analysis of the \AFLOW.org enthalpies \cite{Oses_CHULL_JCIM_2018} indicates that the ``\textit{gain in formation-enthalpy by adding mixing species, $\Delta H_{\rm f}(N+1)-\Delta H_{\rm f}(N)$, decreases with $N$}'',   %CO181009
  In fact, a statistical analysis of the \AFLOW.org enthalpies \cite{Oses_CHULL_JCIM_2018} indicates that the {gain in formation-enthalpy by adding mixing species, $\Delta H_{\rm f}(N+1)-\Delta H_{\rm f}(N)$}, decreases with $N$, 
  and can easily be overcome by the monotonic increase in entropy-gain for the disordered systems (to go from order to complete or partial disorder). 
  In the \AFLOW\ analysis, the threshold between low- and high-entropy systems is around 4 mixing species. To have completely entropy-stabilized materials, 5 mixing-species are required (similar to the Entropy Stabilized Oxides \cite{curtarolo:art99}).
  For carbides in which only the metal-sublattice is randomly populated, 5 metals should be enough to achieve carbide entropy stabilization, especially at equi-composition. 
  Notably, if other sublattices were also allowed to contain disorder, (\textit{e.g.} reciprocal systems like Ta$_x$Hf$_{1-x}$C$_{1-y}$ \cite{Agte_Carbides_ZtP_1930, Andrievskii_Carbides_PM_1967}),  then the overall number of species might reduce. 
  In this example, entropy was increased by introducing point defects (vacancies) --- a promising strategy to improve high temperature performance. 
  In HfC$_{1-x}$ the reduction of C to sub-stoichiometry enhances the stabilizing effect of the configurational entropy on the solid phase, 
  offsetting its Gibbs free energy (vacancies can only exist in the solid phase) leading to an overall increase in melting point (see Ref. \onlinecite{Hong_HfTaC_PRB_2015}). 
}

\noindent
{\bf Experimental results.} To validate the predictions, the {9} chosen carbides are experimentally synthesized and characterized (see Methods section).
The chemical homogeneity of each sample is measured using 
\underline{e}nergy-\underline{d}ispersive x-ray \underline{s}pectroscopy (EDS), 
while the crystalline structure is determined via \underline{x}-\underline{r}ay \underline{d}iffraction (\XRD).

An example of the evolution of a sample of composition HfNbTaTiVC$_5$ through each processing step is given in Figure \ref{fig2}(a), 
demonstrating the densification and homogenization into a single rock-salt structure. At least three distinct precursor phases are 
distinguishable in the mixed powder pattern. Following ball milling, the individual phases are still present, however the peaks 
are considerably broadened, which is due to particle size reduction and mechanical alloying. Following the final \underline{s}park \underline{p}lasma \underline{s}intering (\SPS) step at 
$2200^\circ$C, the sample consolidates into a bulk solid pellet of the desired single rock-salt phase indicating the successful synthesis 
of a high-entropy homogeneous carbide.

Results of \XRD\ analysis for each sample following \SPS\ at $2200^\circ$C, presented in Figure \ref{fig1}(b),
demonstrate that compositions {MoNbTaVWC$_5$, HfNbTaTiZrC$_5$, 
  HfNbTaTiVC$_5$, HfNbTaTiWC$_5$, NbTaTiVWC$_5$, and HfTaTiWZrC$_5$ (the top 6) only exhibit single \FCC\ peaks of 
  the desired high-entropy phase (rock-salt)}, while HfMoTaWZrC$_5$, HfMoTiWZrC$_5$, and HfMoVWZrC$_5$ (the bottom 3) show multiple structures. 
{The small peaks at $2\theta = 31.7^{\circ}$, marked by the diamond symbol in the spectra of HfNbTaTiZrC$_5$, HfNbTaTiWC$_5$ and HfTaTiWZrC$_5$ 
  in Figure \ref{fig1}(b), are from the (111) plane of a monoclinic (Hf,Zr)O$_2$ phase that remains due to processing.
  The volume fraction of this phase is less than 5\% and does not significantly alter the composition of the carbide phase.}
The distinguishable second phase in HfMoTaWZrC$_5$ is identified as a hexagonal phase. 
One and two secondary \FCC\ phases are observed for HfMoVWZrC$_5$ and HfMoTiWZrC$_5$, respectively.
Microstructure analysis and selected elemental mapping (Figure \ref{fig2}(c-f)) confirm that the systems displaying single-phases are chemically 
homogeneous, while the multi-phase samples undergo chemical segregation. 
For example, only grain orientation 
contrast is present in the HfNbTaTiZrC$_5$ microstructure, and no
indication of notable clustering or segregation is visible in its compositional
maps. 
On the contrary, a clear chemical phase contrast is observable in the microstructure of
the multi-phase HfMoTaWZrC$_5$ sample, and the
compositional maps demonstrate that the secondary phase, apparent in
\XRD, is W- and Mo-rich.

{\small
  \begin{table*}
    \caption{\small Results for mechanical properties (bulk: $B$,
      shear: $G$, and elastic moduli: $E$, and Vickers hardness: $H_{\rm V}$) for 6 single-phase high-entropy carbides. 
      Three different models: Chen {\it et al.} \cite{Chen_hardness_Intermetallics_2011}, Teter \cite{Teter_Hardness_MRS_1998}, and Tian {\it et al.} \cite{Tian_IJRMHM_Hardness_2012} 
      are used to calculate theoretical hardness values (only for the 2 non-W-containing compositions) from $B$ and $G$.  
      The \ROM\ values are obtained from the results in this work for rock-salt structure binary carbides, as listed in Table \ref{table3}. 
      Since MoC and WC do not form a stable rock-salt phase at room temperature, 
      {the experimental {\small ROM}s for  Mo- and
        W-containing compositions are not estimated, as indicated by (-)}. 
      Units: $B$, $G$, $E$, and $H_{\rm V}$ in (GPa).}    
    \label{table2}
    % \begin{tabular}{||c||c|c|c||c|c|c||c|c||c|c|c||}
    \footnotesize
    \begin{tabular}{||c||c|c||c|c||c|c||c||c||c||c||}
      \hline
      &		\multicolumn{2}{c||}{$B$} 		&		\multicolumn{2}{c||}{$G$} 		&		\multicolumn{2}{c||}{$E$} 						&		\multicolumn{1}{c||}{$H_{\rm v, Chen}$} 		&		\multicolumn{1}{c||}{$H_{\rm v, Teter}$} 		&		\multicolumn{1}{c||}{$H_{\rm v, Tian}$} 		&		\multicolumn{1}{c||}{$H_{\rm v, exp}$} 		\\  \hline
      system	& \begin{tabular}{@{}c@{}}\fAFLOW \\ (\fROM)\end{tabular}	& \begin{tabular}{@{}c@{}}exp. \\ (\fROM)\end{tabular}&\begin{tabular}{@{}c@{}}\fAFLOW \\ (\fROM)\end{tabular}	&\begin{tabular}{@{}c@{}}exp. \\(\fROM) \end{tabular}&\begin{tabular}{@{}c@{}}\fAFLOW \\ (\fROM)\end{tabular}&\begin{tabular}{@{}c@{}}exp. \\ (\fROM)\end{tabular}&\begin{tabular}{@{}c@{}}\fAFLOW \\ (\fROM)\end{tabular}&\begin{tabular}{@{}c@{}}\fAFLOW \\ (\fROM)\end{tabular}&\begin{tabular}{@{}c@{}}\fAFLOW \\ (\fROM)\end{tabular}&\begin{tabular}{@{}c@{}}exp. \\ (\fROM)\end{tabular}	\\  \hline	
      MoNbTaVWC$_5$	&\begin{tabular}{@{}c@{}}312 \\ (321)\end{tabular}&\begin{tabular}{@{}c@{}} 278 \\(-) \end{tabular}&\begin{tabular}{@{}c@{}}183 \\ (183)\end{tabular}	&\begin{tabular}{@{}c@{}} 226\\(-) \end{tabular}&\begin{tabular}{@{}c@{}}460 \\ (459)\end{tabular}&\begin{tabular}{@{}c@{}} 533$\pm$32\\(-) \end{tabular}	&\begin{tabular}{@{}c@{}}20 \\ (20)\end{tabular}&\begin{tabular}{@{}c@{}}28 \\ (28)\end{tabular}&\begin{tabular}{@{}c@{}}20 \\ (20)\end{tabular}&\begin{tabular}{@{}c@{}} 27$\pm$3\\ (-)\end{tabular}		\\  \hline
      HfNbTaTiZrC$_5$	&\begin{tabular}{@{}c@{}}262 \\ (267)\end{tabular}&\begin{tabular}{@{}c@{}}235 \\ (232)\end{tabular}&\begin{tabular}{@{}c@{}}192 \\ (165)\end{tabular}&\begin{tabular}{@{}c@{}}188 \\(184) \end{tabular}&\begin{tabular}{@{}c@{}}464 \\ (455)\end{tabular}&\begin{tabular}{@{}c@{}}443$\pm$40 \\ (436$\pm$30)\end{tabular}&\begin{tabular}{@{}c@{}}27 \\ (25)\end{tabular}&\begin{tabular}{@{}c@{}}29 \\ (28)\end{tabular}&\begin{tabular}{@{}c@{}}27 \\ (25)\end{tabular}&\begin{tabular}{@{}c@{}}32$\pm$2 \\ (23$\pm$2)\end{tabular}		\\  \hline
      HfNbTaTiVC$_5$	&\begin{tabular}{@{}c@{}}276 \\ (279)\end{tabular}&\begin{tabular}{@{}c@{}} 267 \\ (239)\end{tabular}&\begin{tabular}{@{}c@{}}196 \\ (196)\end{tabular}	&\begin{tabular}{@{}c@{}}212 \\(189) \end{tabular}&\begin{tabular}{@{}c@{}}475 \\ (476)\end{tabular}&\begin{tabular}{@{}c@{}}503$\pm$40 \\ (449$\pm$30)\end{tabular}	&\begin{tabular}{@{}c@{}}26 \\ (26)\end{tabular}&\begin{tabular}{@{}c@{}}30 \\ (30)\end{tabular}&\begin{tabular}{@{}c@{}}26 \\ (26)\end{tabular}&\begin{tabular}{@{}c@{}}29$\pm$3 \\ (24$\pm$2)\end{tabular}	    \\  \hline
      HfNbTaTiWC$_5$	&\begin{tabular}{@{}c@{}}291 \\ (296)\end{tabular}&\begin{tabular}{@{}c@{}} 252 \\ (-)\end{tabular}&\begin{tabular}{@{}c@{}}203 \\ (186)\end{tabular}	&\begin{tabular}{@{}c@{}} 205\\(-) \end{tabular}&\begin{tabular}{@{}c@{}}493 \\ (459)\end{tabular}&\begin{tabular}{@{}c@{}}483$\pm$24 \\(-) \end{tabular}	&\begin{tabular}{@{}c@{}}26 \\ (22)\end{tabular}&\begin{tabular}{@{}c@{}}31 \\ (28)\end{tabular}&\begin{tabular}{@{}c@{}}26 \\ (22)\end{tabular}&\begin{tabular}{@{}c@{}} 31$\pm$2\\(-) \end{tabular}		\\  \hline
      NbTaTiVWC$_5$	&\begin{tabular}{@{}c@{}}305 \\ (304)\end{tabular}&\begin{tabular}{@{}c@{}} 253 \\ (-)\end{tabular}&\begin{tabular}{@{}c@{}}199 \\ (189)\end{tabular}	&\begin{tabular}{@{}c@{}} 206\\(-) \end{tabular}&\begin{tabular}{@{}c@{}}490 \\ (460)\end{tabular}&\begin{tabular}{@{}c@{}}485$\pm$36 \\(-) \end{tabular}	&\begin{tabular}{@{}c@{}}24 \\ (22)\end{tabular}&\begin{tabular}{@{}c@{}}30 \\ (29)\end{tabular}&\begin{tabular}{@{}c@{}}24 \\ (23)\end{tabular}&\begin{tabular}{@{}c@{}} 28$\pm$2\\(-) \end{tabular}		\\  \hline       
      HfTaTiWZrC$_5$	&\begin{tabular}{@{}c@{}}274 \\ (280)\end{tabular}&\begin{tabular}{@{}c@{}} 246 \\(-)\end{tabular}&\begin{tabular}{@{}c@{}}191 \\ (178)\end{tabular}	&\begin{tabular}{@{}c@{}} 200\\(-) \end{tabular}&\begin{tabular}{@{}c@{}}466 \\ (438)\end{tabular}&\begin{tabular}{@{}c@{}}473$\pm$26 \\(-) \end{tabular}	&\begin{tabular}{@{}c@{}}25 \\ (21)\end{tabular}&\begin{tabular}{@{}c@{}}29 \\ (27)\end{tabular}&\begin{tabular}{@{}c@{}}25 \\ (21)\end{tabular}&\begin{tabular}{@{}c@{}} 33$\pm$2\\(-) \end{tabular}		\\  \hline  
    \end{tabular}
  \end{table*}
}

\noindent
{\bf Homogeneity analysis.}
Peak broadening in \XRD\ patterns (Figure \ref{fig1}(b)) is used to quantify the level of structural homogenization achieved in the samples.
% For the 5 compositions exhibiting only rock-salt, but with peaks of different widths, an \underline{analysis} of peak broadening can shed light on the extent of chemical segregation within the single crystal structure. 
According to the Williamson-Hall formulation \cite{Williamson_XRD_Broadening_AM_1953}, broadening in \XRD\ is principally
due to crystallite size ($\propto 1/\cos\theta$, $\theta=$ Bragg angle) and lattice strain ($\propto 1/\tan\theta$).
For multi-component systems, significant broadening is expected to occur due to local lattice
strains and variations in the interplanar-spacings throughout
the sample \cite{Freudenberger_Metals_2017}. The latter can be 
attributed to the inhomogeneous distribution of the elements, the extent of
which can be evaluated by applying the analysis to a
multi-component system, which has only a single lattice structure
and is assumed strain-free \cite{Freudenberger_Metals_2017}.

The lattice distortion of the rock-salt phase in the single-phase materials (or the most prevalent in the multi-phase ones)
is determined by using the relationship between broadening $\beta_{\mathrm S}$ and Bragg angle
$\theta$ (Methods section):
\begin{equation}
  \beta_{\mathrm S} \cos \theta = 4 \varepsilon \sin \theta + \frac{K \lambda}{D}, 
  \vspace{-2mm}
  \label{sample_broadening}
\end{equation}
where $\varepsilon$ is the lattice strain or variation in interplanar-spacing due to chemical inhomogeneity, $K$ is a constant (dependent on the grain shape), $\lambda$ is the incident x-ray 
wavelength, and $D$ is the crystallite size. 
Since materials are assumed strain free, $\varepsilon$ --- obtained by inverting Equation (\ref{sample_broadening}) --- represents the relative 
variation of the lattice parameter due to inhomogeneity.
Thus, $\varepsilon$ is both a measure of homogeneity and of the effective mixing with respect to the ideal scenario.
{The results for the \EFA\ descriptor,
  the experimental characterization,
  as well as the values for $\varepsilon$ for all 9 carbides compositions are given in Table \ref{table1}.
  The values for $\varepsilon$ range from 0.063\% for MoNbTaVWC$_{5}$ and 0.094\% for HfNbTaTiZrC$_5$ (the most homogeneous
  materials) to 0.325\% for HfMoVWZrC$_5$ (the least homogeneous material). 
  Overall, the experimental findings agree well with the predictions of
  \EFA\ descriptor, 
  validate its {\it ansatz} and indicate a potential threshold for our model of 5-metal carbides:
  \EFA\ $\sim 50~\mathrm{(eV/atom)}^{-1} \Rightarrow$} homogeneous disordered single-phase (high-entropy).

\noindent
{\bf High-entropy synthesizability.}
The comparison between the \EFA\ predictions and the homogeneity of the samples is analyzed in Figure \ref{fig2}(b).
Although the Williamson-Hall formalism does not provide particularly accurate absolute values of $\varepsilon$, 
it is effective for comparing similarly processed samples, determining the relative homogeneity.
The lattice distortion $\varepsilon$ (capturing homogeneity) {decreases linearly with the increase of \EFA.
  % The reason is interpreted in the following manner. The \EFA\ is aimed to roughly estimate the distribution of states between energy $E$ and $E+\delta E$.
  % It is similar to the number of states $\Omega$ at a given energy leading to the Boltzmann entropy formula $S = k_{\mathrm B} \log_{\mathrm{e}}\Omega$.
  % Therefore, it is natural to expect logarithmic dependence on the \EFA\ (even without proper normalization) for quantities such as homogeneity ($\varepsilon$).
  The Pearson (linear) correlation of \EFA{}$^{-1}$ %\footnote{The normalization constant, required convert the \EFA\ to a dimensionless quantity so that the logarithm can be calculated, cancels in the expression for the Pearson correlation.}
  with $\varepsilon$ is 0.97, while the Spearman (rank order) correlation is 0.98.
  % As such, the \EFA\ takes the role of an effective ``{\it high-entropy synthesizability descriptor}''.} %CO181009
  As such, the \EFA\ takes the role of an effective {high-entropy synthesizability descriptor}.}

% \cleardoublepage

Three facts are relevant.
{\bf i.} Intuitively, several of the carbide compositions that easily form a highly homogeneous phase, particularly HfNbTaTiVC$_5$ 
and HfNbTaTiZrC$_5$, come from binary precursors having the same structure and ratio of anions to cations as the final high-entropy material.
{\bf ii.} Counterintuitively, the highest-\EFA\ and most homogeneous phase MoNbTaVWC$_{5}$
is made with two precursors having different structures and stoichiometric ratios from the high-entropy material,
specifically orthorhombic $\alpha$-Mo$_2$C and hexagonal $\alpha$-W$_2$C, leading to a final sub-stoichiometric MoNbTaVWC$_{5-x}$.
The additional disorder provided by the presence of vacancies in the C-sublattice is advantageous:
it allows further entropy stabilization, potentially increasing the melting point vis-\`{a}-vis the stoichiometric composition \cite{Hong_HfTaC_PRB_2015}.
An investigation of the effect of carbon stoichiometry is clearly warranted, although it is outside of the scope of this study.
{\bf iii.}
For tungsten and molybdenum, metal-rich carbides are used because of the difficulties in obtaining molybdenum monocarbide (MoC) powder, or tungsten monocarbide (WC) powder in the particle sizes compatible with the other precursors, hindering consistent mixing, sintering and homogenization.
It should be noted that additional samples of MoNbTaVWC$_{5}$ were also synthesized using hexagonal WC with a smaller particle size, 
and the homogeneous rock-salt phase was again successfully obtained (see {Supplementary Figure 37}). 
The existence of phase-pure MoNbTaVWC$_{5}$ indicates that, similar to the rock-salt binary carbides, the multi-component carbide is stable over a range of stoichiometry. 
From experimental/phenomenological grounds, the formation of such a phase is surprising.
The equi-composition binary carbides MoC and WC have hexagonal  ground-states, and their rock-salt configurations have significantly higher formation enthalpy \cite{curtarolo:art75}. 
Considering these facts, there are no experimental indications that adding Mo and W would contribute to stabilizing the most homogeneous rock-salt 5-metal carbide that was predicted by the \EFA\ and later validated experimentally.
{The arguments demonstrate the advantage of a descriptor that quantifies the relative entropy forming ability over simple empirical/phenomenological rules,
  in that it correctly identifies this composition as having a high propensity to form a single phase, while simultaneously correctly predicting that several other systems having both Mo and W 
  subcomponents undergo phase separation.}

\noindent
{\bf Mechanical properties.}
The Vickers hardness, $H_{\mathrm V}$, and elastic modulus, $E$, of
both the binary carbide precursors and the synthesized 5-metal single-phase compositions are measured using nanoindentation, 
where the properties are extracted from load-displacement curves {(see Supplementary Figure 38(a))}.
% as depicted in Figure \ref{fig3}(a).  % FIGURE 3
{The binary precursor samples for these measurements are prepared and analyzed using the same protocol to ensure the validity of the comparisons (see Methods section for more details)}. 
The results for the 5-metal compositions are in Table \ref{table2}, while those for the binary carbides are listed in Table \ref{table3} (Methods section).
It is found that, for the 5-metal compositions, the measured $H_{\mathrm V}$ and $E$ values exceed those predicted from a
\underline{r}ule \underline{o}f \underline{m}ixtures (\ROM) based on the binary precursor measurements. 
% (Figure \ref{fig3}(b)).  % FIGURE 3
The enhancement of the mechanical properties is particularly strong in the case of HfNbTaTiZrC$_5$,
where the measured $E$ and $H_{\mathrm V}$ exceed the \ROM\ predictions by 10\% and  50\%, respectively 
{(see {Supplementary Figure 38(b)} for a comparison between the $H_{\mathrm V}$ results obtained from calculation, experiment, and \ROM)}. 
{Mass disorder is one possible source of the enhanced hardness:
  deformation is caused by dislocation movements and activation energy is absorbed and released at each lattice step.}
An ideal ordered system can be seen as a dislocation-wave-guide with matched (uniform) impedance along the path: 
propagation occurs without any relevant energy reflection and/or dispersion.
This is not the case for disordered systems: mass inhomogeneity causes impedance mismatch, 
generating reflections and disturbing the transmission by dispersing (scattering) its group energy.
Macroscopically the effect is seen as increased resistance to plastic deformations --- more mechanical work is required ---
\textit{i.e.} increase of hardness.
{Other possible causes of increased hardness include solid solution hardening \cite{Ye_hea_high_melting_point,Senkov_HEA_AM_2013,Castle_2018_4metalC}, 
  where the atomic size mismatch results in lattice distortions, limiting the motion of dislocations necessary for plastic deformation; 
  and changes in the slip systems and the ease with which slip can occur \cite{Castle_2018_4metalC, Smith_HfTaC_AM_2018}.} 

Elastic properties are calculated using \AFLOW\ \cite{curtarolo:art110,curtarolo:art115}
% AFLOW to produce a thermally weighted average of the configurations 
for the 5-metal compositions {MoNbTaVWC$_5$,} HfNbTaTiZrC$_5$, HfNbTaTiVC$_5$,
{HfNbTaTiWC$_5$, NbTaTiVWC$_5$, and HfTaTiWZrC$_5$} (Table \ref{table2}) and their precursors.
In general, results are within the experimentally reported ranges for the binary carbides (Table \ref{table3} in the Methods section). 
$H_{\mathrm V}$ values are estimated from the bulk and shear moduli using the models introduced by Chen {\it et al.} \cite{Chen_hardness_Intermetallics_2011}, 
Teter \cite{Teter_Hardness_MRS_1998}, and Tian {\it et al.} \cite{Tian_IJRMHM_Hardness_2012}. 
Computational models do not consider plastic deformation mechanisms in inhomogeneous systems and thus $H_{\mathrm V}$ 
predictions underestimate experiments, leading to results consistent with the \ROM\ of binary carbides (Table \ref{table2}).
The outcome further corroborates that the experimentally observed enhancement of the mechanical properties is due to disorder.

\noindent
{{\bf Vibrational contribution to formation free energy.}
  The vibrational contributions to the formation Gibbs free energy, $\Delta F_{\rm vib}$, at 2000K are listed in Table \ref{table1} for the 6 compositions synthesized as a single phase. 
  The vibrational free energies, $F_{\rm vib}$, are calculated using the Debye-Gr{\"u}neisen model %\cite{Poirier_Earth_Interior_2000} %, BlancoGIBBS2004} 
  implemented in the \AFLOW-\AGL\ module \cite{curtarolo:art96}, using the Poisson ratio calculated with \AFLOW-\AEL\ \cite{curtarolo:art115}. 
  The average $F_{\rm vib}$ for the 5-metal compositions are calculated, weighted according to the Boltzmann distribution at 2000K. 
  The vibrational contribution to the formation Gibbs free energy, $\Delta F_{\rm vib}$, for each composition is obtained from the difference between its average $F_{\rm vib}$ 
  and the average $F_{\rm vib}$ of its component binary carbides.
  $\Delta F_{\rm vib}$ at 2000K ranges from $\sim0$meV/atom for HfNbTaTiWC$_5$ to $-31$meV/atom for HfNbTaTiVC$_5$, 
  which are significantly less than the total entropy contribution (mostly configurational plus vibrational) 
  required to overcome the values of 50meV/atom to 150meV/atom for the formation enthalpy $\Delta H_{\rm f}$. 
  These results are in agreement with previous observations that the vibrational formation entropy is generally an order of magnitude smaller than the configuration entropy \cite{Gao_HEA_Opinion_2017, Axel_RMP}.}

\vspace{5mm}

{\section*{Discussion}}

In this article, an entropy forming ability descriptor has been developed for the purpose of capturing synthesizability of high-entropy materials.
% The framework has been applied to refractory metal carbides, leading to the prediction and subsequent experimental discovery of novel homogeneous high-entropy single-phases.  %CO181009
The framework has been applied to refractory metal carbides, leading to the prediction and subsequent experimental discovery {of homogeneous} high-entropy single-phases.
% A convenient {\it recursive} implementation allows rapid computational prescreening of candidates ($\chi$-prescreening).
% The method is able to pinpoint systems forming homogeneous single-phases and those that do not, thus identifying candidates ripe for experimental synthesis. 
{The method is able to quantitatively predict the relative propensity of each composition to form a homogeneous single-phase, thus identifying the most promising candidates for experimental synthesis.} 
{In particular, the experiments validate the prediction that the composition MoNbTaVWC$_5$ should have a very high propensity to form a homogeneous single-phase, 
  despite incorporating both Mo$_2$C and W$_2$C, which have different
  structures (hexagonal and/or orthorhombic instead of rock-salt) and stoichiometric ratios from the 5-metal high-entropy material.}

Furthermore, it is demonstrated that disorder enhances the mechanical properties of these materials: HfNbTaTiZrC$_{5}$ and
HfTaTiWZrC$_{5}$ are measured to have hardness of 32 GPa (almost 50\% higher than the \ROM\ prediction) and 33 GPa, respectively, 
suggesting a new avenue for designing superhard materials.
% The formalism could become the long-sought enabler of accelerated design for novel high-entropy functional materials   %CO181009
The formalism could become the long-sought enabler of accelerated design {for high-entropy} functional materials  
with enhanced properties for a wide range of different technological applications.

\section*{Methods}
{\small   
  {\small
    \begin{table*}
      \caption{\small Results for mechanical properties (bulk: $B$, shear: $G$, and elastic: $E$, moduli, and Vickers hardness: $H_{\rm V}$) for 8 rock-salt structure binary carbides. 
        The \AFLOW\ values are calculated using the Voigt-Reuss-Hill average and the \underline{A}utomatic \underline{E}lasticity \underline{L}ibrary (\AEL) module \cite{curtarolo:art115}, 
        while $H_{\rm V}$ is estimated using 3 different models described in the literature. 
        These results are compared with two sets of available measured data, obtained from the literature \cite{C_and_N_mech_prop_book_1989,The_chem_of_TM_C_and_N_book_1996} 
        and the current experiments. Units: $B$, $G$, $E$, and $H_{\rm V}$ in (GPa). 
        $^{\rm a}$ This work. 
        $^{\rm b}$ $\alpha$-MoC$_{1-x}$ + $\eta$-MoC + $\gamma$-MoC composite \cite {Dubitsky_2007_rsMoC_hardness}. 
        $^{\rm c}$ WC$_{1-x}$; $x$=0.36-0.41 \cite {Caron_2011_rsWC_hardness}.
      }      
      \label{table3}
      % \begin{tabular}{||c||c|c|c||c|c|c||c|c|c|c|c||}
      \footnotesize
      \begin{tabular}{||c||c|c|c||c|c|c||c|c|c||c|c|c|c|c||}
        \hline
        &		\multicolumn{3}{c||}{$B$} 	&		\multicolumn{3}{c||}{$G$} 	&		\multicolumn{3}{c||}{$E$} 	&		\multicolumn{5}{c||}{$H_{\rm V}$}					\\  \hline									
        system	&	\fAFLOW\	&	 exp. &exp.$^{\rm a}$	&	\fAFLOW\	&	 exp.  &exp.$^{\rm a}$	&	\fAFLOW\	&	 exp. 	&	exp.$^{\rm a}$	&	Chen	&	Teter	&	Tian	&	exp. 	&	exp.$^{\rm a}$	\\  \hline
        HfC	&	239	&	241	&223	&186	&	179-193	&181	&443	&	316-461	&	428$\pm$32	&	29	&	28	&	28	&	19-25	&	25$\pm$2	\\  \hline
        MoC	&	335	&	-	&-	&152	&	-	&-	&396	&	-	&	-	&	12	&	23	&	13	&27-83$^{\rm b}$	&-		\\  \hline
        NbC	&	297	&	296-378	&246	&199	&	197-245	&177	&488	&	330-537	&	429$\pm$46	&	25	&	30	&	25	&	19-25	&	17$\pm$3	\\  \hline
        TaC	&	326	&	248-343	&219	&213	&	215-227	&184	&525	&	241-722	&	431$\pm$44	&	25	&	32	&	25	&	16-23	&	14$\pm$2	\\  \hline
        TiC	&	251	&	241	&	255	&181	&186	&207&438	&	447-451	&	489$\pm$13	&	26	&	27	&	25	&	32	&	31$\pm$2	\\  \hline	
        VC	&	283	&	389	&	250	&199	&157	&196	&484	&	268-420	&	465$\pm$13	&	26	&	30	&	26	&	20-29	&	29$\pm$1	\\  \hline
        WC	&	365	&	-	&-	&153	&	-	&-	&403	&	-	&	-	&	11	&	23	&	12	&	$>$28$^{\rm c}$	&-		\\  \hline
        ZrC	&	221	&	220	&	216	&157	&172	&169	&381	&	385-406	&	402$\pm$13	&	23	&	28	&	22	&	23-25	&	24$\pm$1	\\  \hline
      \end{tabular} 
    \end{table*}
  }
  
  \noindent
  \textbf{Spectrum generation.} The different possible configurations required to calculate the energy spectrum are generated using the \AFLOWPOCC\ algorithm \cite{curtarolo:art110} 
  implemented within the \AFLOW\ computational materials design framework \cite{curtarolo:art65, curtarolo:art75, curtarolo:art104}. 
  The algorithm initially generates a superlattice of the minimum size necessary to obtain the required partial occupancies within some user-specified accuracy. For each 
  unique superlattice, the \AFLOWPOCC\ algorithm then generates the complete set of possible supercells using Hermite normal form matrices \cite{curtarolo:art110}. %, gus_enum1}. 
  Non-unique supercell combinations are eliminated from the ensemble by first estimating the total energies of all configurations using a 
  Universal Force Field \cite{curtarolo:art110, Rappe_1992_JCAS_UFF} based method, and then identifying duplicates from their identical energies. 
  % Structural comparisons are then performed between configurations which are found to have the same energy, and any duplicates are removed from the calculation set. 
  \\
  \noindent\textbf{Structure generation.} 
  In the case of the high-entropy carbide \cite{curtarolo:art99} systems investigated here, the \AFLOWPOCC\ algorithm starts with the rock-salt crystal structure 
  (spacegroup: $Fm\overline{3}m,\ \#225$; Pearson symbol: cF8; \AFLOW\ Prototype: {\sf AB\_cF8\_225\_a\_b} \cite{curtarolo:art121}) as the input parent lattice. 
  Each anion site is occupied with a C atom (occupancy probability of 1.0), while the cation site is occupied by 5 different refractory metal elements, with a 0.2 occupancy probability for each. 
  The \AFLOWPOCC\ algorithm then generates a set of configurations (49 in total in the case of the rock-salt based 5-metal carbide systems, once structural duplicates are excluded), each containing 10 atoms:
  one atom of each of the metals, along with 5 carbon atoms. 
  This is the minimum cell size necessary to accurately reproduce the required stoichiometry. 
  All configurations have $g_{i} = 10$, except for one where $g_{i} = 120$, so that $\sum_{i=1}^{n}g_{i} = 600$ for the rock-salt based 5-metal carbide systems.
  {Note that computational demands increase significantly with the number of elements: \AFLOWPOCC\ generates 522, 1793 and 7483 for 6-, 7- and 8-metal carbide compositions, respectively.}  
  \\
  \noindent \textbf{Energies calculation.} 
  The energy of each configuration is calculated using density functional theory ({\small VASP} \cite{vasp_prb1996}) within the 
  \AFLOW\ framework \cite{curtarolo:art65} and the standard settings \cite{curtarolo:art104}. 
  Each configuration is fully relaxed using the {\small PBE} parameterization of the {\small GGA} exchange-correlation functional \cite{PBE},
  {\small PAW} potentials,  at least 8000 {\bf k}-points per reciprocal atom ({\small KPPRA}),  and a plane-wave cut-off of at least 1.4 times the cut-off values of constituent species' pseudopotentials \cite{curtarolo:art104}. 
  The formation enthalpy ({$H_{\rm f}$}) of each configuration along
  with the link to \AFLOW{}.org entry page is provided in
  {Supplementary Tables 4-10}.

  \noindent
  \textbf{Mechanical properties.}  Elastic properties are calculated using the \underline{A}utomatic \underline{E}lasticity \underline{L}ibrary (\AEL) module \cite{curtarolo:art115} 
  of the \AFLOW\ framework, which applies a set of independent directional normal and shear strains to the structure, and fits the resulting stress tensors to obtain the elastic constants.
  From this, the bulk: $B$, and shear: $G$, moduli are calculated in the Voigt, Reuss and Voigt-Reuss-Hill (\VRH) approximations, 
  with the average being used for the purposes of this work. 
  {The elastic or Young's modulus: $E$, is calculated using the approximation $E =9BG/(3B+G)$, which can be derived starting from the expression for
    Hooke's Law in terms of $E$ and the Poisson ratio, $\nu$: $\epsilon_{11} = 1/E \left[\sigma_{11} - \nu \left(\sigma_{22} + \sigma_{33} \right) \right]$ \cite{suh_mechanical}, and similarly for $\epsilon_{22}$ and $\epsilon_{33}$. 
    For a cubic system, $\epsilon_{11} = S_{11} \sigma_{11} + S_{12} \sigma_{22} + S_{12} \sigma_{33}$ (similarly for $\epsilon_{22}$ and $\epsilon_{33}$), where $S_{ij}$ are the elements of the elastic compliance tensor, 
    so that $1/E = S_{11}$ and $-\nu/E = S_{12}$. 
    For a cubic system, the bulk modulus is $B = 1/[3(S_{11} + 2 S_{12})] = E/[3(1-2\nu)]$.
    The Poisson ratio can be written as $\nu = (3B-2G)/(6B+2G)$, and combining with the expression for $B$ and rearranging gives the required $E =9BG/(3B+G)$.}

  The elastic properties for the 5-metal compositions are first calculated for each of the 49 configurations generated by \AFLOWPOCC. 
  {The \VRH\ approximated values of $B$ and $G$ for these
    configurations are listed in {Supplementary Table
      3}, along with the \AFLOWPOCC\ ensemble averaged electronic
    density of states (see {Supplementary Figure 39}).} 
  The average elastic moduli are then obtained, weighted according to the Boltzmann distribution at a temperature of 2200$^\circ$C (the experimental sintering temperature). 
  These calculated values are compared to those obtained using a \underline{r}ule \underline{o}f \underline{m}ixtures (ROM, average of the binary components, weighted according to fractional composition in the sample). 

  Three different models are used for predicting the Vickers hardness based on the elastic moduli: 
  Chen {\it et al.} ($H_{\mathrm V} = 2(k^2G)^{0.585}-3$; $k =G/B$) \cite{Chen_hardness_Intermetallics_2011}, 
  Teter ($H_{\mathrm V} = 0.151G$) \cite{Teter_Hardness_MRS_1998}, 
  and Tian {\it et al.} ($H_{\mathrm V} = 0.92k^{1.137}G^{0.708}$; $k =G/B$) \cite{Tian_IJRMHM_Hardness_2012}. 
  Note, however, that these models are based on the elastic response of the materials, and do not take into account phenomena such as plastic deformation, slip planes, and lattice defects.

  \noindent
  {\bf Sample preparation.} 
  Initial powders of each of the eight binary precursor carbides (HfC, NbC, TaC, TiC, Mo$_2$C, VC, W$_2$C, ZrC) are obtained in greater than 99\%
  purity and $-325$ mesh ($<44 \mu$m) particle size (Alfa Aesar). Samples are weighed out in 15 g batches and mixed to achieve the desired 5-metal
  carbide compositions. To ensure adequate mixing, each sample is ball milled in a shaker pot mill for a total of 2 hours in individual
  30-minute intervals intersected by 10-minute rest times to avoid heating and consequent oxide formation. All milling is done in
  tungsten carbide lined stainless steel milling jars with tungsten carbide grinding media.

  Bulk sample pellets are synthesized via solid-state processing routes. The \underline{f}ield \underline{a}ssisted \underline{s}intering \underline{t}echnique ({\small FAST}), 
  also called \underline{s}park \underline{p}lasma \underline{s}intering (\SPS), 
  is employed to simultaneously densify and react the compositions into single-phase materials. For all samples, sintering is done at $2200^\circ$C with a heating 
  rate of $100^\circ$C/min, 30 MPa uniaxial pressure, and a 5-minute dwell at temperature. Samples are heated in vacuum atmosphere to $1300^\circ$C followed by flowing 
  argon to $2200^{\circ}$C. All sintering is done in 20 mm graphite die and plunger sets with graphite foil surrounding the samples to prevent reaction with the die.

  \noindent
  {\bf Sample analysis.}
  Elemental analysis is performed using an FEI Quanta 600 SEM equipped with a Bruker e-Flash EDS
  detector at an accelerating voltage of 20 kV. 
  Microstructural SEM imaging is carried out using an FEI Apreo FE-SEM at an accelerating voltage of 5 kV, 
  with a combination of secondary and back-scattered electron detectors to show phase contrast.
  Crystal phase analysis is performed using a Rigaku Miniflex X-ray
  Diffractometer with a stepsize of $0.02^\circ$ and 5 second dwells, using Cu K$\alpha$ radiation 
  (wavelength $\lambda = 1.54059$\AA) for all measurements and calculation of the lattice parameter.
  All sample patterns are fitted in Materials Data Incorporated's
  ({\small MDI}) Jade 9 software \cite{jade9} with a residual of fit $R < 8\%$. 
  Lattice parameter, $a_{\mathrm{exp}}$, values of {4.353\AA, 4.500\AA, 4.415\AA,
    4.434\AA, 4.355\AA, 4.502\AA, 4.506\AA, 4.534\AA, and 4.476\AA\ were measured for MoNbTaVWC$_5$, HfNbTaTiZrC$_5$,
    HfNbTaTiVC$_5$, HfNbTaTiWC$_5$, NbTaTiVWC$_5$, HfTaTiWZrC$_5$, HfMoTaWZrC$_5$, HfMoVWZrC$_5$, and HfMoTiWZrC$_5$}, respectively 
  (for multi-phase samples, $a_{\mathrm{exp}}$ refers to the primary cubic phase).

  For analysis of sample peak broadening $\beta_{\mathrm S}$, instrumental broadening $\beta_{\mathrm I}$ must first be determined. For this, a NIST 660b LaB$_6$ standard is run under the 
  same conditions as each carbide sample. The instrumental profile is then fitted, and $\beta_{\mathrm I}$ is determined to vary with Bragg angle $\theta$ as:
  \begin{equation}
    \beta_{\mathrm I} = 0.1750985 - 0.001560626 \theta + 0.00001125342 \theta^2.
    \vspace{-1mm}
    \label{instrument_broadening}
  \end{equation}

  $\beta_{\mathrm S}$ is determined by subtracting $\beta_{\mathrm I}$ from the measured broadening $\beta_{\mathrm M}$: $\beta^x_{\mathrm S} = \beta^x_{\mathrm M} - \beta^x_{\mathrm I}$.
  $\beta_{\mathrm M}$ is measured as a function of $\theta$, and $x$ is a constant between 1.0 and 2.0. In the current analysis, $x$ is set to 2.0 due to the Gaussian-like shape of the 
  instrument peaks, as this value leads to the lowest standard deviation of linear fits to the peak broadening data.

  Both crystallite size and lattice strain contribute to $\beta_{\mathrm S}$ 
  \cite{Williamson_XRD_Broadening_AM_1953, Cullity_XRD_book_1956, Mote_XRD_ZnO_2012}:
  \begin{equation}
    \beta_{\mathrm S} = 4 \varepsilon \tan \theta + \frac{K \lambda}{D \cos \theta},
    % \vspace{-2mm}
    \label{sample_broadening_methods}
  \end{equation}
  where $\varepsilon$ is the lattice strain or variation in interplanar spacing due to chemical inhomogeneity, $K$ is a constant (dependent on the grain shape), $\lambda$ is the incident x-ray 
  wavelength and $D$ is the crystallite size. 
  Rearranging Equation (\ref{sample_broadening_methods}) gives:
  \begin{equation}
    \beta_{\mathrm S} \cos \theta = 4 \varepsilon \sin \theta + \frac{K \lambda}{D}.
    \vspace{-2mm}
    \label{sample_broadening_cosine}
  \end{equation}
  The slope of a linear fit to the plot of $\beta_{\mathrm S} \cos \theta$ against $\sin \theta$ is equal to the strain, or lattice distortion, while the $y$-intercept 
  of a linear fit with zero slope determines the crystallite size.

  \noindent
  {\bf Mechanical Testing.}
  Mechanical properties of each of the single-phase compositions are tested using a Keysight NanoIndenter G200 with a Berkovich indenter tip. 
  To rule out indentation size effects, testing is carried out at 
  loads of both 50 mN and 300 mN, and no significant deviation in hardness or modulus is observed. 
  To allow valid cross-comparison, each of the HECs is compared to the
  binary carbides, which were hot-pressed and indentation tested under
  identical conditions. For the reported values, tests are carried out
  according to the standard method outlined in ISO 14577 using a
  maximum load of 50mN. Values are calculated as an average of 40 indents, and are reported with errors of plus or minus one standard deviation. 
  A fused crystal silica standard is run prior to each test to ensure proper equipment calibration is maintained. 
  Samples are polycrystalline with grain sizes between 10$\mu$m and 30$\mu$m. 
  Prior to indentation testing each sample is vibratory polished using 0.05$\mu$m colloidal silica for 12 hours to ensure minimal surface roughness.  
  All tests are carried out at a temperature of $27^{\circ}$C $\pm$ $0.5^{\circ}$C. 
  Indentation data is analyzed according to the methods of Oliver and Pharr \cite{oliver_pharr_1992, oliver_pharr_2004}.
  The elastic (\textit{i.e.} Young's) modulus is determined using $1/E_{\mathrm{eff}} = (1 - \nu^2)/E + (1 - \nu_{\mathrm I}^2)/E_{\mathrm I} $, 
  where $E_{\mathrm{eff}}$ is the effective modulus (sometimes called the reduced modulus) obtained from nano-indentation, 
  $E$ and $\nu$ are the Young's modulus and Poisson's ratio, respectively, for the specimen, while $E_{\mathrm I}$ and $\nu_{\mathrm I}$ are the same parameters for the indenter.
  A Poisson's ratio for each of the binary carbides is obtained from literature \cite{Kral_Lengauer_1998} where available. 
  For 5-metal carbide samples where data for each of the constituents is available, the value used for Poisson's ratio is taken as the average of the constituent binaries. 
  If this average is unavailable ({\it i.e.} when Mo and/or W are present), Poisson's ratio is assumed to be equal to 0.18. 
  
  \noindent
  {\bf Data availability.}
  All the \textit{ab-initio} data are freely available to the public as
  part of the \AFLOW\ online repository and can be accessed through {\small AFLOW}.org
  following the {\small REST-API} interface \cite{curtarolo:art65} and {\small AFLUX} search language \cite{curtarolo:art128}.
}

\section*{Acknowledgments}
{\small
  \noindent
  The authors acknowledge support 
  by DOD-ONR (N00014-15-1-2863, N00014-17-1-2090, N00014-16-1-2583, N00014-17-1-2876)
  and by Duke University -- Center for Materials Genomics -- for computational support.
  S.C. acknowledges the Alexander von Humboldt Foundation for financial support.
  C.O. acknowledges support from the National Science Foundation Graduate Research Fellowship under Grant No. DGF1106401.
  The authors thank Axel van de Walle, Matthias Scheffler, Claudia Draxl, Ohad Levy, Yoav Lederer, Amir Natan, Omar Cedillos Barraza, Joshua Gild, Olivia Dippo, and Cameron McElfresh for helpful discussions.
}

\bibliographystyle{PhysRevwithTitles_noDOI_v1b.bst}
\newcommand{\Ozolins}{Ozoli\c{n}\v{s}}

\end{document}